\documentclass[numberedappendix]{openjournal}

\usepackage{lipsum}

\usepackage{xcolor}
\usepackage{textgreek}
\usepackage[utf8]{inputenc}
\usepackage[english]{babel}

\usepackage{hyperref}
\hypersetup{
    unicode, 
    colorlinks=true,
    linkcolor=linkcolor,
    citecolor=linkcolor,
    filecolor=linkcolor,
    urlcolor=linkcolor,
}
\usepackage{color,colortbl}
\definecolor{linkcolor}{rgb}{0.0,0.3,0.5}
\usepackage{tensind}
\tensordelimiter{?}
\DeclareGraphicsExtensions{.bmp,.png,.jpg,.pdf}
\usepackage{verbatim}
\usepackage[normalem]{ulem}
\usepackage{orcidlink}
\usepackage{soul}
\usepackage{booktabs}

\graphicspath{ {./figs/} }

\usepackage{amsmath, amssymb}
\usepackage{bm}
\usepackage{natbib}
\usepackage{cancel}
\newcommand{\bmu}{{\boldsymbol{\mu}}}
\newcommand{\btheta}{{\boldsymbol{\theta}}}
\newcommand{\bt}{\boldsymbol{t}}
\newcommand{\bx}{\boldsymbol{x}}

\newcommand{\bC}{\boldsymbol{C}}
\newcommand{\bD}{\boldsymbol{D}}
\newcommand{\bM}{\boldsymbol{M}}
\newcommand{\bN}{\boldsymbol{N}}
\newcommand{\bQ}{\boldsymbol{Q}}

\newcommand{\bS}{\boldsymbol{S}}
\newcommand{\bU}{\boldsymbol{U}}
\newcommand{\bW}{\boldsymbol{W}}
\newcommand{\bX}{\boldsymbol{X}}
\newcommand{\bY}{\boldsymbol{Y}}
\newcommand{\bZ}{\boldsymbol{Z}}
\newcommand{\bLambda}{\boldsymbol{\Lambda}}
\newcommand{\bSigma}{\boldsymbol{\Sigma}}
\newcommand{\QuadForm}[1]{\bt^\top #1 \bt}

\begin{document}

\title{On combining estimated and analytic covariance matrices}
\author{Alan Heavens\orcidlink{0000-0003-1586-2773}}
\email{a.heavens@imperial.ac.uk}
\affiliation{Imperial Centre for Inference and Cosmology (ICIC), Department of Physics, Imperial College, London SW7 2AZ, UK}
\author{Lorne Whiteway\orcidlink{0000-0001-6375-0755}}
\email{lorne.whiteway.13@ucl.ac.uk}
\affiliation{Department of Physics \& Astronomy, University College London, Gower Street, London, WC1E 6BT, UK}
\author{Elena Sellentin\orcidlink{0009-0002-2655-3458. }}
\email{sellentin@strw.leidenuniv.nl}
\affiliation{Mathematical Institute and Leiden Observatory, Leiden University, Gorlaeus Gebouw, Einsteinweg 55, NL-2333 CC Leiden, The Netherlands}
\begin{abstract}
The statistical analysis of cosmological data often assumes a Gaussian sampling distribution and relies on covariance matrices estimated from simulations. In this setting, the likelihood function of the data is not Gaussian but is instead a multivariate Student-$t$ distribution, arising from marginalisation over an inverse-Wishart distribution for the true covariance matrix. This framework, introduced by \citet{SH2016} and extended by \citet{Percival2022}, provides a principled drop-in replacement to the Gaussian likelihood with Hartlap correction \citep{Hartlap2007}. The latter removes bias in the precision matrix; it is still widely used, despite failing to reproduce the heavy tails of the true distribution (thus yielding inaccurate probabilities, especially in the case of tensions between datasets).

In practice, cosmological analyses frequently involve additional Gaussian error contributions, for example from instrumental noise, foregrounds, super-sample covariance, or emulator uncertainties. The resulting likelihood function is a convolution of the Sellentin-Heavens or Percival et al. likelihoods with an extra Gaussian contribution, and does not have a simple expression.  In this
note, we derive an accurate approximation for the combined likelihood function, another multivariate Student-$t$ distribution which inherits the heavy tails.  The parameters of the Student-$t$ distribution are determined by matching the covariance and multivariate kurtosis to those of the true distribution. We identify many application areas where this may be useful, and derive here simple-to-apply practical formulae for the effective degrees of freedom and scale matrix of the Student-$t$ likelihood function when some element of the covariance matrix is estimated from simulations.

We also include a slightly more expensive but fast sampling algorithm, based on the mixture representation of the Student-$t$ distribution, which avoids the approximation altogether, but is not the drop-in replacement for the normal Gaussian or Hartlap likelihood function that the Student-$t$ approximation in this paper provides.

\end{abstract}

\section{Introduction}
The analysis of cosmological data has traditionally involved the construction of informative summary statistics, such as estimates of the clustering correlation function or power spectrum.  It is then common to assume a Gaussian sampling distribution (or likelihood function).  While there are more sophisticated techniques, such as field-level inference \citep[e.g.][]{JascheWandelt,JascheLavaux,Porqueres21} or simulation-based inference \citep[e.g.][]{Alsing,Makinen24,Jeffrey}, summary statistics with assumed Gaussian distributions remain the typical choice for the first analysis of new datasets.  Often the assumptions are not questioned, despite the highly non-Gaussian nature of present-day cosmological fields, and the Gaussian likelihood approximation may be poor, even in cases where the central limit theorem might suggest otherwise \citep[e.g.][]{SellentinHeavens2018}.  

Accepting the Gaussian premise also involves further challenges, notably through the difficulty of providing a covariance matrix (a four-point function).  Often the only way this can be done is via estimation from a large number of simulations.  It is the inverse of the covariance matrix that appears in the Gaussian likelihood function, and it is well-known that the inverse of the estimated covariance matrix is biased.  An approach widely used in cosmology has been to correct for this bias using the Hartlap factor \citep{Hartlap2007}. This method is straightforward. It was adopted in many early survey pipelines, and it continues to be widely used, despite not giving the correct likelihood, in particular being unable to capture the heavy tails induced by covariance uncertainty. 
In contrast, marginalising analytically over the inverse-Wishart posterior for the covariance leads exactly to a multivariate Student-$t$ likelihood \citep{SH2016}; this can be adapted to a frequency-matching prior to give good frequentist coverage \citep{Percival2022}.  Alternative approaches have included a shrinkage estimator to combine some prior knowledge of the covariance matrix \citep{HallTaylor}.

The aim of this work is to generalise this Student-$t$ framework to the following fairly common setting: one element of variability has a covariance that is estimated from simulations, and there is another source of Gaussian variability that has an analytic or otherwise theoretically computable covariance.   This setting may arise for example in cosmic shear studies, where an analytic super-sample covariance \citep[e.g.][]{Linke, Schreiner} is added to a within-sample covariance estimated from simulations.  The situation can also arise in Bayesian hierarchical models (BHMs), which are increasingly applied in cosmology to capture the propagation of uncertainties across multiple levels of data modelling.  Marginalisation over uncertain covariances within one level of a BHM naturally gives rise to a multivariate Student-$t$ contribution to the likelihood. 

In addition, extra Gaussian components commonly enter the hierarchy, for example from shape noise \citep[e.g.][]{Barreira, Upham} or from emulator uncertainties. The convolution of Student-$t$ and Gaussian distributions therefore emerges naturally within BHMs, but it lacks a simple closed-form expression.  In super-sample covariance, fluctuations of background modes larger than the survey window modulate small-scale power, producing a coherent covariance component that can be computed analytically or semi-analytically via response functions and the survey window \citep{TakadaHu2013,LiHuTakada2014}. Such a combination is common in weak lensing and galaxy clustering studies \citep[e.g.][]{BarreiraSchmidt2017,Joachimi2021}.    In another application area, the computational expense of running cosmological simulations has spurred the adoption of emulators using Gaussian processes or neural networks.  Uncertainty in the emulator predictions contribute an additional,  analytic prediction variance that depends on training design and kernel hyperparameters \citep[e.g.][]{Heitmann2014,Rogers2019,Percival2022}.   In the Cosmic Microwave Background at low multipoles the distribution of estimated angular power spectra is non-Gaussian, often modelled by Wishart or Student-$t$ forms \citep{HamimecheLewis2008}. At the same time, instrumental noise is commonly assumed to be Gaussian.   

Given the prevalence in cosmology of the combination of Student-$t$ and Gaussian distributions, it is useful to have an accurate approximation for the resulting likelihood function,  retaining the heavy-tailed behaviour while keeping the likelihood analytically tractable. In classical statistics, the convolution of independent distributions and the matching of moments is a well-studied problem \citep{Kotz2004}; such an approach can be applied to this problem, so that we approximate the convolution by another Student-$t$ distribution, with modified parameters set by matching the covariance as well as the multivariate kurtosis of \citet{Mardia1970}.  We provide here compact expressions for the resultant likelihood function.  The approximation is inevitably imperfect (Appendix \ref{app:nonelliptical}). 

As an alternative, we can define an efficient but approximation-free sampler using the mixture representation of the Student-$t$ distribution in the convolution.  We use this both to demonstrate the accuracy of the Student-$t$ approximation, and as a standalone algorithm for sampling the exact distribution.

\section{Covariance Marginalisation}
Before we consider the combination of Student-$t$ and Gaussian-distributed variable, we review briefly the derivations of the Student-$t$ distributions that arise from uncertainty in the true covariance matrix $\bC$, when only an estimate from simulations is known.  

Let us assume that we have a data vector drawn from a Gaussian distribution of length $p$ and mean $\bmu$: $\bX \sim {\cal N}(\bmu,\bC)$.  An estimate of the covariance matrix $\hat{\bC}$ is obtained from $N_{\rm sim}$ simulations:
\begin{equation}
    \hat{\bC} \equiv \bS = \frac{1}{N_{\rm sim}-1}\sum_{i=1}^{N_{\rm sim}}(\bX_i-\bar\bX_i)(\bX_i-\bar\bX_i)^\top,
\end{equation}
where $\bar\bX$ is the sample mean.
The sample covariance matrix follows a Wishart distribution \citep{Anderson2003}:
\begin{equation}
(N_{\rm sim}-1){\bS} \sim \mathcal{W}_p(\bC,\,N_{\rm sim}-1).
\end{equation}
Given $\bS$, the true covariance matrix $\bC$ is uncertain, so to construct the likelihood for $\bX$ we marginalise over it:
\begin{equation}
    P(\bX|\bS) \propto \int\,P(\bX|\bC)\,P(\bS|\bC)\,P(\bC)\,d\bC.
\end{equation}
\citet{SH2016} chose to marginalise $\bC$ with a Jeffreys prior \citep{YangBerger1994}
$P(\bC) \propto |\bC|^{-(p+1)/2}$ for the true covariance, deriving the exact likelihood for the $p$-dimensional data $\bX$ conditioned on $\bS$:
\begin{equation}
  \mathcal{L}_{\mathrm{SH}}(\bX|\btheta,\bS) \propto |\bS|^{-1/2}
  \left(1 + \frac{\chi_S^2}{N_{\mathrm{sim}}-1}\right)^{-\frac{N_{\mathrm{sim}}}{2}};
  \quad \chi_S^2 = (\bX-\bmu)^\top \bS^{-1} (\bX-\bmu),
  \label{eq:SH0}
\end{equation}
where $\btheta$ are the model parameters, which are usually to be inferred.

\citet{Percival2022} modified this by introducing a frequency-matching prior, $P(\bC) \propto |\bC|^{-(m-N_{\mathrm{sim}}+p+1)/2}$, which then modifies the likelihood of the data to
\begin{equation}
  \mathcal{L}_{\mathrm{Percival}}(\bX|\btheta,\bS) \propto |\bS|^{-1/2}
  \left(1 + \frac{\chi_S^2}{N_{\mathrm{sim}}-1}\right)^{-m/2},
  \label{eq:Percival0}
\end{equation}
with
\begin{equation}
  m = 2 + N_\btheta + \frac{N_{\mathrm{sim}} - 1 + B(p - N_\btheta)}
                                {1 + B(p - N_\btheta)}, \quad
  B = \frac{N_{\mathrm{sim}}-p-2}{(N_{\mathrm{sim}}-p-1)(N_{\mathrm{sim}}-p-4)},
\end{equation}
and where $N_\btheta$ is the number of parameters in the problem.  
This distribution allows the posteriors to be interpreted as frequentist confidence intervals. It reduces to the Sellentin-Heavens likelihood in the large-$N_{\mathrm{sim}}$ limit.   

\subsection{Sellentin-Heavens and Percival et al. likelihoods as Student-$t$ distributions}

The Student-$t$ distribution is defined, for $p$-dimensional data $\bX$ with $\nu$ degrees of freedom, location vector $\boldsymbol{\mu}$, and 
scale matrix $\bSigma$, by 
\begin{equation}
\bX \sim t_\nu(\bmu,\bSigma) \;\propto\; 
|\bSigma|^{-1/2} \left[ 1 + \frac{1}{\nu} 
  (\bX-\boldsymbol{\mu})^\top 
  \bSigma^{-1} (\bX-\boldsymbol{\mu}) 
\right]^{-(\nu+p)/2}.
\end{equation}
The likelihood functions (\ref{eq:SH0} and \ref{eq:Percival0}), although not obviously Student-$t$ distributions, can nevertheless be put in this form. The \citet{SH2016} likelihood (\ref{eq:SH0}) has parameters:
\begin{equation}
    \nu_{\mathrm SH} = N_{\rm sim}-p; \quad \bSigma_{{\mathrm SH}} = \frac{N_{\rm sim}-1}{N_{\rm sim}-p}\,\bS,
    \label{eq:SH}
\end{equation}
while the frequency-matched \cite{Percival2022} likelihood (\ref{eq:Percival0}) has parameters:
\begin{equation}
 \nu_{\mathrm P} = m-p; \quad \bSigma_{\mathrm P} = \frac{N_{\rm sim}-1}{m-p}\,\bS.
 \label{eq:Percival}
\end{equation}
The requirement $\nu = m-p > 4$ (for the kurtosis to be well-defined) requires more (typically $\sim 2 p$) simulations than are needed simply for $\bS$ to be invertible.  If $m<p+4$ then the function exists and is everywhere positive, but it is not a Student-$t$ distribution, as this would require $\nu>1$.  The function is not a proper sampling distribution in this case, as its integral over the data space diverges and so it cannot be normalised.  One might be tempted to use it as an improper pseudo-likelihood distribution; however, the approximation in this paper cannot be used, since it relies on matching the covariance and kurtosis, both of which diverge.  Whether another approximation is possible, perhaps based on matching curvature at the peak, is not something that we have pursued.

\section{Approximate distribution of the sum of Student-$t$ and Gaussian distributed variables}

Let us consider the situation in which simulations are used to estimate one component of variability, but another non-simulated additive component (such as super-sample covariance, from implicit separation of scales \citep{TakadaHu2013} or measurement error) needs to be included. We assume all elements have underlying Gaussian distributions.  

Consider a Student-$t$ distributed variable $\bX$ with degrees of freedom $\nu$ and scale matrix $\mathbf{\Sigma}$:  $\bX\sim t_\nu(\mathbf{0},\mathbf{\Sigma})$, and an independent Gaussian-distributed variable $\bY$ with covariance matrix $\bSigma_g$: $\bY \sim \mathcal{N}(\mathbf{0},\bSigma_g)$.  For simplicity we assume both distributions have zero mean.   

The distribution of $\bZ \equiv \bX + \bY$ is a convolution with no simple expression, but with known moments.  It inherits the heavy tails of the Student-$t$ distribution, motivating its approximation by another Student-$t$ distribution with different degrees of freedom $\nu_\star$ and scale matrix $\bSigma_\star$, i.e. we approximate
\begin{equation}
    \bZ \; \dot{\sim} \; t_{\nu_\star}(\mathbf{0},\bSigma_\star).
\end{equation}
The goal is to find $\nu_{*}$ and $\bSigma_\star$; we do this by matching the covariance and multivariate kurtosis to those of the true distribution.  

The covariance of the original Student-$t$ distribution is $\nu \bSigma/(\nu-2)$ for $\nu>2$, and, since $\bX$ and $\bY$ are independent, the total covariance is the sum
\begin{equation}
\bSigma_Z = \bSigma_g + \left(\frac{\nu}{\nu-2} \right)\bSigma.
\label{eq:totalcov}
\end{equation}
In addition to matching the covariance, we equate the \citet{Mardia1970} multivariate kurtosis of $\bZ$,
$\mathbb{E}\left[ (\bZ^\top \bSigma_Z^{-1} \bZ)^2 \right]$, to the Mardia kurtosis of a Student-$t$ distribution with degrees of freedom $\nu_\star$,
\begin{equation}
{\rm Kurt(t_{\nu_\star})} = \frac{p(p+2)(\nu_\star-2)}{\nu_\star-4}.
\end{equation}

The full derivation is in appendix \ref{app:kurtosis_matching}, and we simply quote the results here:
\begin{equation}
\nu_\star = 4+\frac{2\,p(p+2)}{\kappa}; \qquad \bSigma_\star = 
\left(\frac{\nu_{*}-2}{\nu_{*}}\right) \bSigma_Z,
\label{eq:keyresults}
\end{equation}
where
\begin{equation}
\kappa=\frac{2\nu^2}{(\nu-2)^2(\nu-4)}\left[
(\mathrm{tr}\,\bN)^2+2\,\mathrm{tr}(\bN^2)\right];\quad \bN = \bSigma_Z^{-1}\bSigma.
\label{eq:kappa}
\end{equation}

The key quantities  $\nu_\star$ and $\bSigma_\star$ (\ref{eq:keyresults}) that define the approximate distribution can be calculated for the addition of Gaussian noise to the Student-$t$ distributions of \citet{SH2016} or \citet{Percival2022} by appropriate insertion of $\nu=\nu_{SH}$ or $\nu_{P}$ and scale matrix $\bSigma=\bSigma_{SH}$ or $\bSigma_P$ from (\ref{eq:SH}) or (\ref{eq:Percival}).  For the new Student-$t$ distribution to be proper, the minimum number of simulations is increased compared to the case of no additional Gaussian component.  This reduces the differences with a Gaussian approximation, but nevertheless is more accurate, especially in the tails; this can be important in the presence of tensions, when tail probabilities can be very different.

\section{Numerical experiments}
\label{sec:experiments}

We illustrate the method with synthetic data generated from 
\begin{equation}
y_i=\theta_0+\theta_1 x_i+\epsilon_i;\ 
x_i\sim {\cal U}(-1,1);\  \epsilon_i \sim {\cal N}(0,\sigma_g^2);\ i=1,\ldots,p,
\end{equation}
with true $\theta_0=0$, $\theta_1=1$. 
A covariance matrix  was drawn from the frequency-matching prior of \citet{Percival2022} and covariance matrices were estimated from data drawn from a zero-mean Gaussian with this covariance.  The Gaussian part of known covariance was included by drawing a zero-mean Gaussian data vector.

To test the accuracy of the approximation, we perform coverage tests, sampling the covariance matrices from the Percival et al. prior. The distribution favours near-singular matrices, so sampling is non-trivial; see appendix \ref{app:fibre_measure} for how this was achieved.

\begin{figure}[h]
\centering
\includegraphics[width=0.7\textwidth]{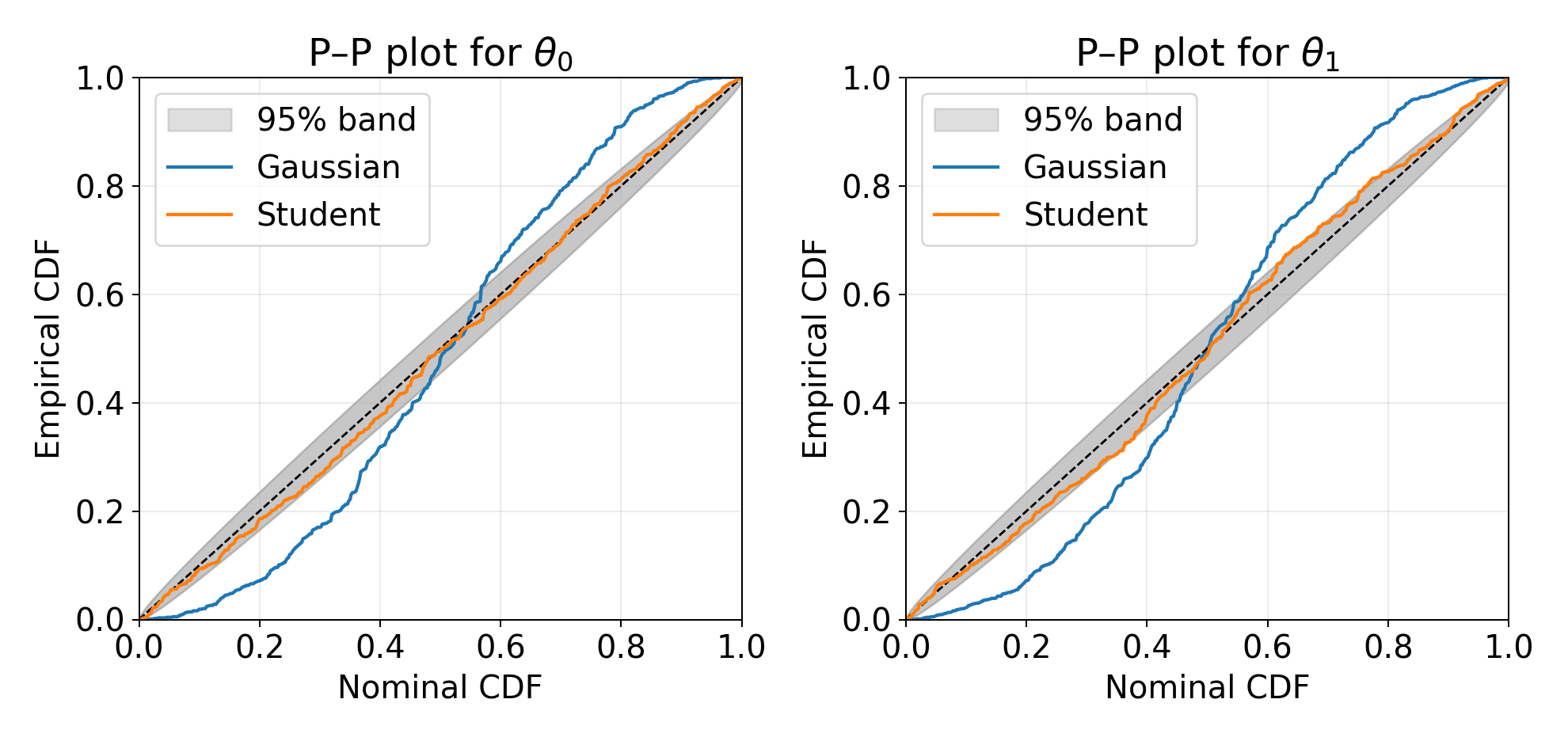}
\caption{Coverage P--P plots for the approximate Student-$t$ distribution (orange).  Also shown in blue is the Gaussian approximation. Data were generated from $y_i=\theta_0+\theta_1 x_i+\epsilon_i$ with $\theta_0=0$,  $\theta_1=1$, and $x_i\sim {\cal U}(-1,1)$, $p=100$ and $N_{\rm sim}=210$.  The Gaussian noise covariance was drawn from a random distribution with the same determinant prior, and multiplied by a strength parameter $\Sigma_g=0.1$. Posterior inference of $(\theta_0,\theta_1)$  was performed in Stan with 100 warmup samples, and 1000 inference samples, across $M=500$  experiments. The dashed line is the ideal 1--1 coverage, and the grey band is expected variation. Covariance matrices were drawn from the frequency-matching prior of \citet{Percival2022}.
}
\label{fig:coverage}
\end{figure}

\begin{figure}[h]
\centering
\includegraphics[width=0.7\textwidth]{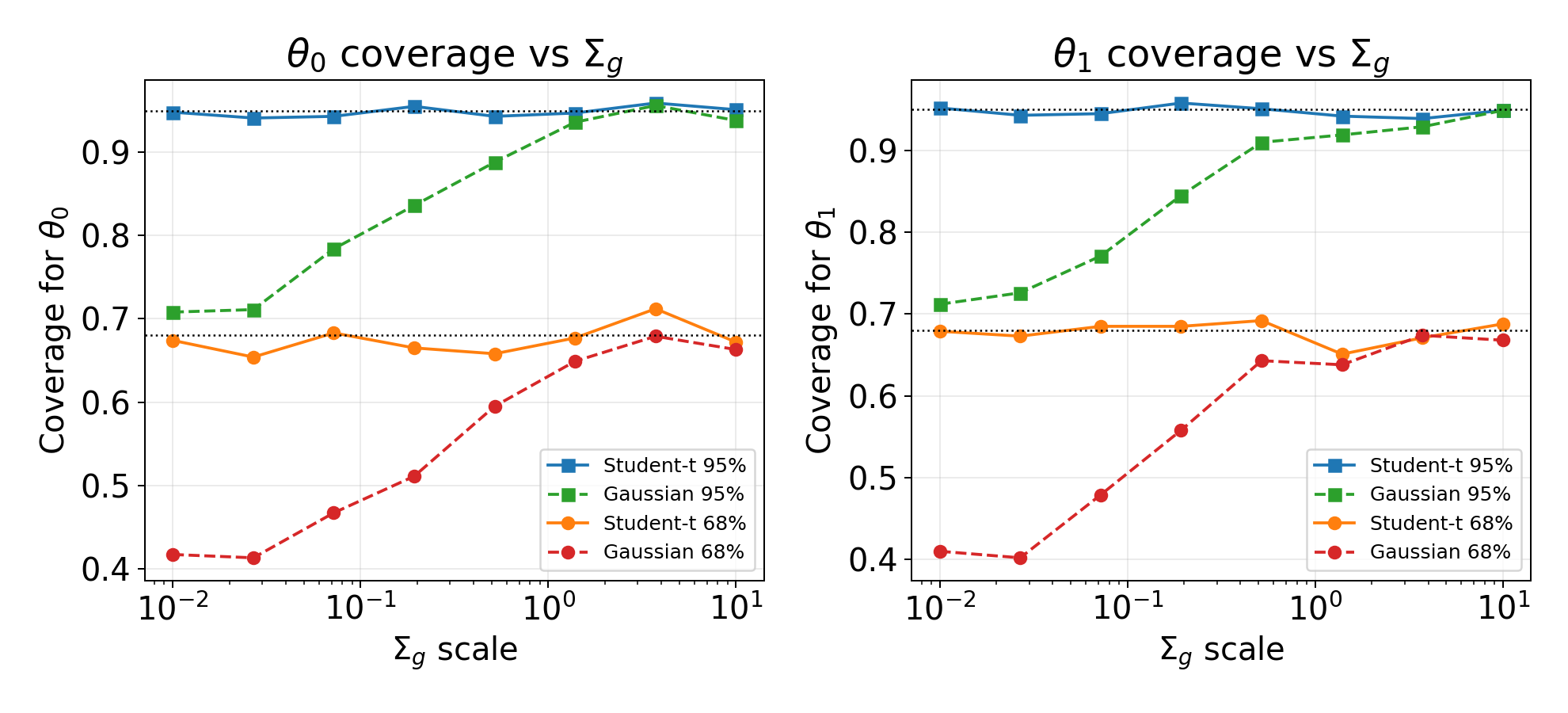}
\caption{68\% and 95\% coverage for varying amplitude of the Gaussian contribution, for the setup shown in Fig.~\ref{fig:coverage}.  The covariance-matched Gaussian approximation fares badly except when the Gaussian contribution is dominant ($\Sigma_g\gg 1$), whereas the approximate Student-$t$ likelihood has excellent coverage for all levels of Gaussian noise. $\Sigma_g$ scale is the multiplier of the noise covariance matrix drawn from the \cite{Percival2022} prior.}
\label{fig:coverage2}
\end{figure}

The sampler package Stan \citep{carpenter2017stan,stan_dev_team} was used with NUTS \citep{hoffman2014nuts} for the inference with typically 100 warmup samples and $10^3-10^4$ inference samples. Fig.~\ref{fig:coverage} shows a coverage P--P plot for a low-noise setup, showing improved coverage of the Student-$t$ over the covariance-matched Gaussian approximation. The determinant of the noise covariance was drawn from the \cite{Percival2022} prior ($p=100$, $N_{\rm sim}=210$), and multiplied by $\Sigma_g = 0.1$. Fig.~\ref{fig:coverage2} shows how well the Student-$t$ 68\% and 95\% credible regions match frequentist coverage for increasing noise $\Sigma_g$, compared with the covariance-matched Gaussian approximation.  As can be seen, the approximation is very accurate over the entire dynamic range of the test.

\section{Tomographic weak lensing}

Here we investigate more realistic cases, namely cosmic shear surveys. One survey is relatively small and the other is large; with each the within-survey covariance is estimated from simulations, and there is an additional analytic Gaussian super-sample covariance component.  CAMB \citep{CAMB} was used to generate power spectra, which were used to train an emulator for speed. The cosmological setup is a flat Universe with $\Omega_m=0.3$, $S_8=0.8$, $\Omega_b=0.049$, $h=0.67$,  $\sum m_\mu = 0.06$ eV. $n_s=0.965$, with a nonlinear power spectrum computed with Halofit \citep{takahashi}. We assume wide uniform priors on $S_8$ and $\Omega_m$. We use OneCovariance \citep{OneCovariance} to generate covariance matrices, combining the Gaussian and non-Gaussian terms into a single within-sample covariance, and generate a super-sample covariance which is assumed known.  We used five tomographic bins, with setups being example configurations provided in OneCovariance with their default $n(z)$ and one-component ellipticity standard deviations, and two survey areas of 777 and 14000 square degrees.  Further details are given in Table \ref{tab:survey_params}. All auto-power spectra and five cross-spectra are used in the analysis, with 15 band powers in $\ell$, giving a total of 150 data points.  The within-sample covariance is estimated from 350 simulations at the fiducial point $S_8 = 0.8,\ \Omega_m=0.3$.  With the \cite{Percival2022} frequency-matching prior, this is specifically chosen to be in a regime where the Student-$t$ corrections are large: the scale matrix is 6.7 times the sample variance $\bS$, and the resulting large uncertainty in the true covariance matrix leads to much larger posteriors than we would expect in a more realistic case with a higher number of simulations. The super-sample covariance is heavily sub-dominant in these cases.  

\begin{table}[!htbp]
\centering
\begin{tabular}{ l c c c c c c c }
\toprule
& Area/deg$^2$ & Median redshifts & Number density/arcmin$^2$ & $\sigma_\epsilon$ & $\ell_{\rm min}$ &$\ell_{\rm max}$ & $\ell$ bins\\
\midrule
Small survey & 777 & 0.33, 0.53, 0.70, 0.90, 1.23 & 1.2, 1.5, 1.8, 1.3, 1.2 & 0.27,0.26,0.28,0.27,0.29  & 20 & 5000 & 15\\
\midrule
Large survey & 14000 & 0.33, 0.53, 0.70, 0.90, 1.23 & 6.0, 7.5, 9.0, 6.5, 6.0 & 0.27,0.26,0.28,0.27,0.29 & 20 & 5000 & 15\\
\bottomrule
\end{tabular}
\caption{The two sets of survey characteristics used to test the method with super-sample covariance as the Gaussian addition.  The survey characteristics are example setups in OneCovariance \citep{OneCovariance}.}
\label{tab:survey_params}
\end{table}

\subsection{Accurate posterior with hierarchical model}
\label{sec:bhm}

We compare three approximations with the exact posterior sampled using Stan from the hierarchical model.  This sampling can be done efficiently by writing the Student-$t$ distribution as a mixture representation, as in section \ref{sec:mixture}.   The full posterior for the set of power spectra $\bx$ is a convolution, given by equations similar to eq. (\ref{eq:mixture}) and eq. (\ref{eq:lambda}), where in these probability distributions we have suppressed some dependencies for clarity:
\begin{equation}
\bx\,|\,\tau \sim 
\mathcal{N}\!\left(\mathbf{\bmu},\,\mathbf{\bSigma_g}+\mathbf{\bSigma}/\tau\right); \qquad
\tau\sim \mathrm{Gamma}\!\left(\frac{\nu}{2},\,\frac{\nu}{2}\right).
\label{eq:convolution}
\end{equation}
$\mathbf{\bmu}(S_8,\Omega_m)$ contains the theoretical band powers.  We sample the three parameters ($S_8,\Omega_m, \tau$) with Stan, and marginalise over $\tau$.  

\subsection{Cosmology results}

\begin{figure}[h]
\centering
\includegraphics[width=0.5\textwidth]{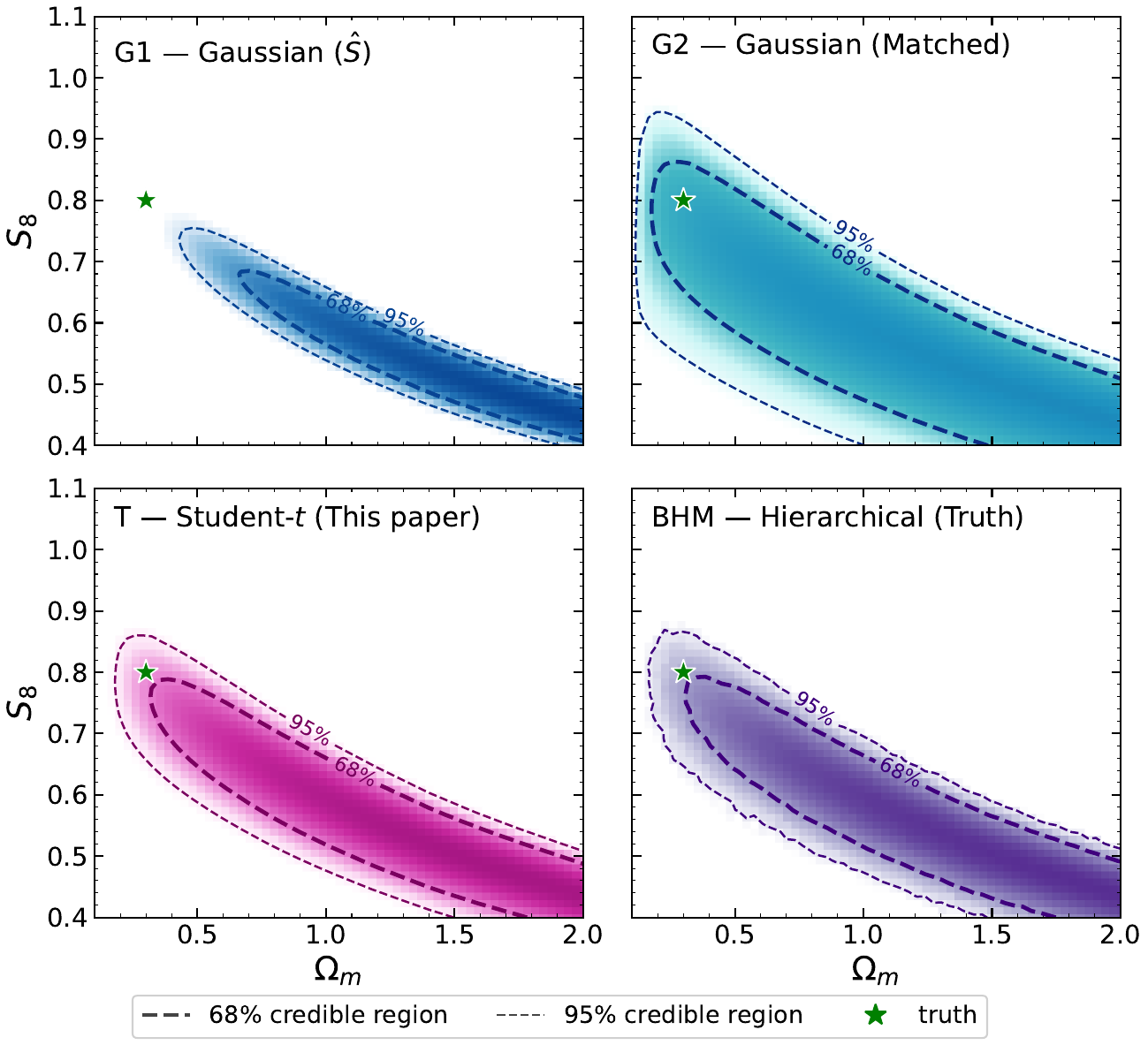}
\caption{Inference of the matter density parameter $\Omega_m$ and the clustering strength $S_8$, from a cosmic shear survey of 777 square degrees and 5 tomographic bins, with Gaussian super-sample covariance. Top left: Posterior using na\"ive Gaussian likelihood function, with total covarance $\bS + \bC_{\rm ssc}$.  Top right: Gaussian likelihood function, with covariance matched to the true convolved value. Bottom left: approximate Student-$t$ likelihood function.  Bottom right: samples of the true posterior using a hierarchical model as described in the text.}
\label{fig:kids}
\end{figure}

\begin{figure}[h]
\centering
\includegraphics[width=0.5\textwidth]{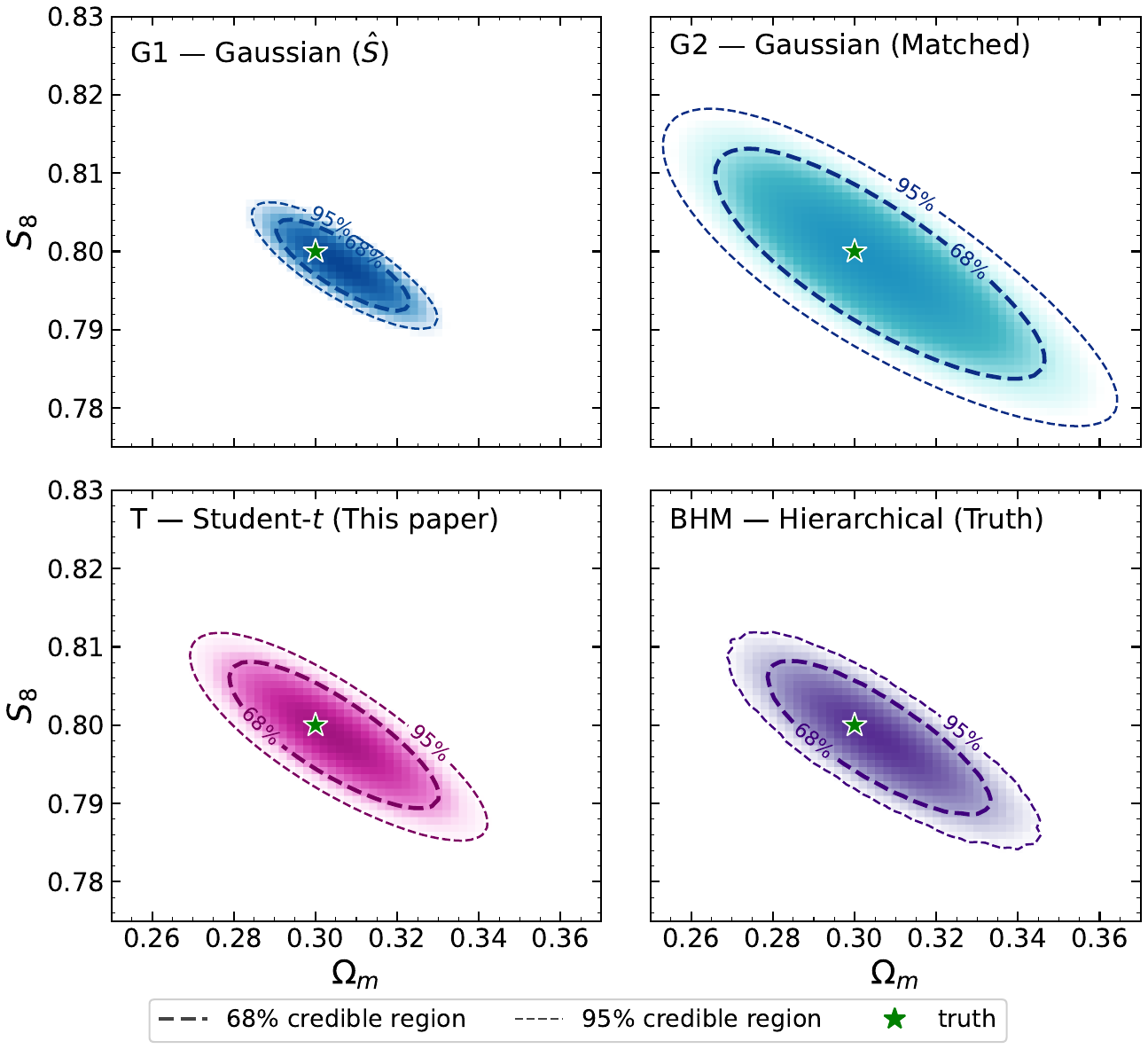}
\caption{As in Fig. \ref{fig:kids}, but with a larger 14000 square degree survey coverage and higher number density.}
\label{fig:euclid}
\end{figure}

In Fig. \ref{fig:kids} and \ref{fig:euclid} we show the resultant posteriors for the two surveys. In each figure, the bottom right panel shows the true posterior, obtained by exact sampling from the hierarchical model. The top left panel shows a Gaussian approximation in which we take the estimated within-sample covariance and add it to the super-sample covariance; this severely underestimates the error, as it does not account for the uncertainty in the true covariance matrix. The top right panel also shows a Gaussian approximation, in this case obtained by matching the total covariance to the true value; this too is a poor approximation to the true posterior. The bottom left panel shows the Student-$t$ approximation, in which we match covariance and kurtosis. Note that the covariance matching is done in the data space - the covariance of the parameters is not guaranteed to match, and does not. We see that the Student-$t$ provides a very good approximation to the full likelihood for these realistic cases, and we suggest that either it or the full hierarchical model needs to be used when the number of simulations is not far in excess of the number of data points.

\section{Conclusions}
\label{sec:conclusions}

We have presented approximate but accurate solutions for the convolution of Gaussian and Student-$t$ distributions, motivated by problems in cosmological data analysis where the covariance matrix of one element of fluctuations is estimated from simulations, and is then combined with another Gaussian source of variability, such as super-sample covariance, shape noise in weak lensing, or measurement error. In one and many dimensions, we showed how to construct an effective multivariate Student-$t$ distribution by matching covariance and kurtosis, following the framework of multivariate kurtosis introduced by \citet{Mardia1970} and subsequent treatments of multivariate $t$-distributions \citep[e.g.][]{Kotz2004}. This approximation retains heavy tails and analytic tractability, while accommodating additional Gaussian contributions such as instrumental noise, foreground residuals, or emulator errors.

Our results generalise the Student-$t$ framework introduced by \citet{SH2016} and extended by \citet{Percival2022}, which arise naturally from marginalising over covariance uncertainty. In contrast, the widely used Gaussian likelihood with Hartlap correction \citep{Hartlap2007} only corrects the bias of the inverse covariance in expectation, while leaving the likelihood Gaussian, and in particular underestimating the sampling distribution in the tails. The approximation developed here provides a simple and principled solution that preserves the heavy-tailed behaviour expected from finite-simulation covariance estimation, and remains valid when further Gaussian noise contributions are present.  The approximation can also be used in Bayesian hierarchical models (BHMs), where uncertainties from multiple levels of the hierarchy propagate into the effective likelihood, and the method can combine levels of the hierarchy for efficiency if needed in a statistically principled way.  

From a practical perspective, these effective Student-$t$ likelihoods are straightforward to implement in cosmological inference pipelines, requiring only adjusted degrees of freedom and scale matrix - it is basically a simple fix to existing Gaussian or Student-$t$ codes.  Specifically, the (natural) log likelihood is given simply by
\begin{equation}
\log{P(\bX|\bS,\bSigma_g)} = {\rm const} - \frac{(\nu_\star+p)}{2}\log\left[1+\frac{1}{\nu_\star}{(\bX-\bmu)^\top\bSigma_\star^{-1}(\bX-\bmu)}\right]
\end{equation}
where $\nu_\star$ and $\bSigma_\star$ are given by eq. (\ref{eq:keyresults}), using eq. (\ref{eq:kappa}), which is obtained from the frequency-matching prior values of $\nu=m-p$ and $\bSigma=\bSigma_P$ given in eq. (\ref{eq:Percival}).  A Python implementation is provided in Appendix \ref{app:code}.

 As an alternative, an efficient sampling of the exact convolution can be done using the mixture representation as shown in section \ref{sec:bhm}.

\section*{Data Availability}
No separate data are used for this article. Reproducibility can be established by generating datasets according to the specification of the model in section \ref{sec:experiments}.

\section*{Acknowledgments}

LLMs were used in an initial literature sweep and in some code generation. 

\appendix

\section{Non-elliptical distribution}
\renewcommand{\theequation}{A.\arabic{equation}}
\setcounter{equation}{0}
\label{app:nonelliptical}

We show that a Student-$t$ approximation to the convolution of a Student-$t$ and a Gaussian is necessarily inexact. Recall that elliptical distributions are precisely those for which the characteristic function has the form 
\begin{equation}
    \phi(\bt) \equiv \mathbb{E}[\exp(i \bt^\top \bX)] = \exp(i \bt^\top \boldsymbol{\mu}) \, \psi\!\left(\QuadForm{\bSigma}\right)
\end{equation}
for some vector $\boldsymbol{\mu}$, nonnegative-definite $\bSigma$, and scalar \emph{}{characteristic generator} function $\psi$. The Student-$t$ and Gaussian distributions are both elliptic: for the Student-$t$ distribution, $\boldsymbol{\mu}$ and $\boldsymbol{\Sigma}$ are the mean and the shape matrix, and the characteristic generator is
\begin{equation}
    \psi_{\textrm{ST}}(u) \propto u^{\nu/4} K_{\nu/2}(\sqrt{\nu u})
\end{equation}
(where $K$ is the modified Bessel function of the second kind), while for the Gaussian distribution, $\boldsymbol{\mu}$ and $\boldsymbol{\Sigma}$ are the mean and the covariance matrix, and the characteristic generator is
\begin{equation}
    \psi_{\textrm{G}}(u) = \exp\left(-\frac{1}{2}u\right) \, .
\end{equation}
Recall also that the characteristic function of the sum of two independent random variables is the product of the individual characteristic functions.

In what follows we assume that all distributions have zero mean; the extension to non-zero means is straightforward. Let $\bX \sim t_\nu(\mathbf{0},\mathbf{\bSigma})$ be a Student-$t$ distribution and $\bY\sim\mathcal{N}(\mathbf{0},\bSigma_g)$ an independent Gaussian. Assume that $\bZ = \bX + \bY$ is Student-$t$ say $\bZ \sim t_\nu(\mathbf{0},\bSigma_\star)$. Taking the natural log of the characteristic generators and rearranging yields
\begin{equation}
    -\frac{1}{2}\QuadForm{\Sigma_g} = \left[\frac{\nu}{4} \log(\QuadForm{\bSigma_\star}) + \log( K_{\nu/2}(\sqrt{\nu \QuadForm{\bSigma_\star}}))\right] - \left[\frac{\nu}{4} \log(\QuadForm{\bSigma}) + \log( K_{\nu/2}(\sqrt{\nu \QuadForm{\bSigma}}))\right] + \textrm{const} \, .
\end{equation}
But here the LHS is a polynomial in $\bt$ while the RHS is not. This contradiction shows that $\bZ$ cannot be Student-$t$, and so our approximation must be inexact.

We can say more; we show that if $\bSigma_g$ and $\bSigma$ are not scalar multiples of each other then $\bZ$ cannot be \emph{any} elliptical distribution i.e. it cannot have elliptical contours. If $\bZ$ is elliptical, then the natural log of its characteristic generator $\psi_\star$ will satisfy
\begin{equation}
    \log\left[\psi_\star(\QuadForm{\bSigma_\star})\right] = \frac{\nu}{4} \log(\QuadForm{\bSigma}) + \log\left[ K_{\nu/2}(\sqrt{\nu \QuadForm{\bSigma}})\right]-\frac{1}{2}\QuadForm{\bSigma_g} + \textrm{const}
    \label{eq:log_char_gen}
\end{equation}
for some $\bSigma_\star$ (which must be positive definite because $\bSigma$ and $\bSigma_g$ are). Let $E$ be the set of $\bt$ for which the quadratic form $\QuadForm{\bSigma_\star}$ is unity. Fix $\bt \in E$, and evaluate eq.  (\ref{eq:log_char_gen}) at $\sqrt{u} \bt$; we obtain
\begin{equation}
    \log\left[\psi_\star(u)\right] = \frac{\nu}{4} \log(u \, \QuadForm{\bSigma}) + \log\left[ K_{\nu/2}(\sqrt{\nu \, u \, \QuadForm{\bSigma}})\right]-\frac{1}{2} \, u \, \QuadForm{\bSigma_g} + \textrm{const} \, .
    \label{eq:log_char_gen_u}
\end{equation}
This fixes the definition of $\psi_\star$. The LHS is independent of $\bt$, so the RHS must be as well. However on the RHS there can be no cancellation between the constant, linear, log, and log Bessel terms. Thus the quadratic forms $\QuadForm{\bSigma}$ and $\QuadForm{\bSigma_g}$ must both be constant on $E$. This can only happen if $\bSigma$ and $\bSigma_g$ are scalar multiples of $\bSigma_\star$ and hence of each other; in this latter case while eq. (\ref{eq:log_char_gen_u}) gives the characteristic function of $\bZ$, the corresponding probability density function will not have a simple closed form.

\section{Covariance and kurtosis matching for the Gaussian-Student-$t$ convolution}
\renewcommand{\theequation}{B.\arabic{equation}}
\setcounter{equation}{0}
\label{app:kurtosis_matching}

We derive the parameters of that multivariate
Student-$t$ distribution that matches both the covariance and 
the Mardia’s multivariate kurtosis of the convolution of a Student-$t$ with a Gaussian. 

Let $\bX \sim t_\nu(\mathbf{0},\mathbf{\bSigma})$ have a $p$-dimensional 
Student-$t$ distribution with $\nu>4$ degrees of freedom and scale matrix $\bSigma$ and let $\bY\sim\mathcal{N}(\mathbf{0},\bSigma_g)$ 
be an independent Gaussian variable.  We will consider $\bZ=\bX+\bY$.

\subsection{Mixture representation}
\label{sec:mixture}

The multivariate Student-$t$ distribution with a scale matrix $\bSigma$ and degrees of freedom $\nu$ can be written as a scale--mixture representation
\citep{AndrewsMallows1974,Kotz2004}, $p(\bX|\bSigma,\nu) = \int\, d\tau\, p(\bX|\bSigma,\tau)\,p(\tau |\nu)$, where, suppressing some dependencies for clarity,
\begin{equation}
\bX\,|\,\tau \sim 
\mathcal{N}\!\left(\mathbf{0},\frac{\mathbf{\bSigma}}{\tau}\right),
\qquad
\tau\sim \mathrm{Gamma}\!\left(\frac{\nu}{2},\,\frac{\nu}{2}\right).
\label{eq:mixture}
\end{equation}
Hence,  conditional on $\tau$,
\begin{equation}
\bZ\,|\,\tau \sim 
\mathcal{N}\!\left(\mathbf{0},\,\mathbf{\bSigma_g}+\frac{\mathbf{\bSigma}}{\tau}\right).
\label{eq:lambda}
\end{equation}
Note that when marginalised over $\tau$, (\ref{eq:lambda}) does not give an exact Student-$t$ distribution for $\bZ$.

The total covariance of $\bZ$ is obtained from integrating $\mathbb{E}[\bZ\bZ^\top | \tau]$ over $\tau$.  Using $\mathbb{E}[1/\tau] = \nu/(\nu-2)$ for a Gamma distribution, we find
\begin{equation}
\bSigma_{\bZ} = \mathbb{E}[\bZ\bZ^\top] 
= \mathbf{\bSigma_g}+\left(\frac{\nu}{\nu-2}\right)\,\mathbf{\bSigma}.
\label{eq:SigmaZ}
\end{equation}

\subsection{Mardia’s kurtosis}

For Gaussian-distributed $\bW \sim \mathcal{N}(\mathbf{0},\bSigma)$ and an arbitrary symmetric matrix $\bM$, we see via Wick's theorem that
\begin{equation}
\mathbb{E}\!\left[\left(\bW^\top \bM\,\bW\right)^2\right]
=\left[\mathrm{tr}(\bM\bSigma)\right]^2 +
2\,\mathrm{tr}\!\left[(\bM\bSigma)^2\right].
\label{eq:quad_form}
\end{equation}

Now for an arbitrary mean-zero distribution $\bD$, the kurtosis of \citet{Mardia1970} is defined to be
\begin{equation}
\textrm{Kurt}(\bD) = \mathbb{E}\left[ (\bU^\top \bSigma_{\bD}^{-1} \bU)^2 \right],
\end{equation}
where $\bSigma_{\bD}$ is the covariance of $\bD$ and $\bU \sim \bD$.
Using eq. (\ref{eq:quad_form}) with $\bM=\bSigma_g^{-1}$ we see that the Mardia kurtosis of a $p$-dimensional Gaussian is $p(p+2)$.
This, together with the mixture representation and with $\mathbb{E}[1/\tau^2]=\nu^2/[(\nu-2)(\nu-4)]$, then shows that the Mardia kurtosis of a $p$-dimensional Student-$t$ distribution with $\nu$ degrees of freedom is
\begin{equation}
{\mathrm{Kurt}}(t_{\nu}) = \frac{p(p+2)(\nu-2)}{\nu-4}.
\end{equation}

\subsection{Expectation over the mixture}

Setting $\bM=\bSigma_{\bZ}^{-1}$ in eq. (\ref{eq:quad_form}) and using the mixture representation, we find
\begin{equation}
{\mathrm{Kurt}}(\bZ)=p(p+2)
+\frac{2\nu^2}{(\nu-2)^2(\nu-4)}\left[
(\mathrm{tr}\,\bN)^2+2\,\mathrm{tr}(\bN^2)\right] \equiv p(p+2) + \kappa,
\label{eq:betaZ}
\end{equation}
where $\bN=\bSigma_{\bf Z}^{-1}\mathbf{\bSigma}$.

We now match the covariance and kurtosis of $\bZ$ to that of a Student-$t$ distribution with degrees of freedom $\nu_\star$ and scale matrix $\bSigma_\star$, and hence covariance $\nu_\star \bSigma_\star/(\nu_\star-2)$ and kurtosis
$p(p+2)(\nu_\star-2)/(\nu_\star-4)$.
This yields the effective degrees of freedom
\begin{equation}
\nu_\star = 4+\frac{2\,p(p+2)}{\kappa}
\label{eq:nustar}
\end{equation}
and scale matrix
\begin{equation}
\bSigma_\star = \frac{(\nu_\star-2)}{\nu_\star}\left[\bSigma_g+\left(\frac{\nu}{\nu-2}\right)\bSigma\right].
\label{eq:Sstar}
\end{equation}
Together, \eqref{eq:nustar} and \eqref{eq:Sstar} provide the parameters of the 
effective Student-$t$ approximation.

\section{Lebesgue measure, eigenvalues, and uniform sampling on determinant fibres}
\label{app:fibre_measure}
\renewcommand{\theequation}{C.\arabic{equation}}
\setcounter{equation}{0}

We derive the probability density for the eigenvalues of a covariance
matrix under a Lebesgue-induced prior of the form $P(\bC)\propto |\bC|^{q}$.

\subsection{Lebesgue measure on symmetric positive-definite matrices}

Let $C$ be a real, symmetric, positive-definite $p\times p$ matrix.  The
Lebesgue measure on the space of symmetric matrices is
\begin{equation}
d\bC = \prod_{i\le j} dC_{ij}.
\end{equation}
Any absolutely continuous prior $P(\bC)$ is defined with respect to this measure.

We consider priors of the form
\begin{equation}
P(\bC)\,d\bC \;\propto\; |\bC|^{q}\, d\bC,
\end{equation}
where $q\in\mathbb{R}$ (with $q=-(p+1)/2$ corresponding to the Jeffreys prior considered by \cite{SH2016}).

\subsection{Eigenvalue--eigenvector decomposition}

Following \citet{YangBerger1994}, we proceed via eigenvalue--eigenvector factorisation of the covariance matrix:
\begin{equation}
\bC = \bQ\,\mathbf{\Lambda}\,\bQ^{\mathsf T},
\qquad
\mathbf{\Lambda} = \mathrm{diag}(\lambda_1,\dots,\lambda_p),
\end{equation}
with $\bQ$ an orthogonal matrix and $\lambda_i>0$.  With this change of variables, the Lebesgue measure factorises as (see \citealt{Muirhead1982} p.105)
\begin{equation}
d\bC
\;=\;
\left(\prod_{i<j}|\lambda_i-\lambda_j|\right)
\left(\prod_{i=1}^p d\lambda_i\right)
d\mu(\bQ),
\label{eq:jacobian_lambda}
\end{equation}
where $d\mu(\bQ)$ is Haar measure on the orthogonal group $\mathrm{O}(p)$.
The factor $\prod_{i<j}|\lambda_i-\lambda_j|$ is the Vandermonde determinant.

Including the determinant weight gives
\begin{equation}
P(\lambda)\,\prod_i d\lambda_i
\;\propto\;
\left(\prod_i \lambda_i^{q}\right)
\left(\prod_{i<j}|\lambda_i-\lambda_j|\right)
\prod_i d\lambda_i.
\end{equation}
Defining the (natural) log-eigenvalues $\ell_i = \log \lambda_i$, the density becomes
\begin{equation}
P(\ell)\,\prod_i d\ell_i
\;\propto\;
\exp\!\left[(q+1)\sum_i \ell_i\right]
\prod_{i<j}\left|\exp(\ell_i)-\exp(\ell_j)\right|
\prod_i d\ell_i.
\end{equation}
Using
\[
|\exp({\ell_i})-\exp({\ell_j})|
= 2 \exp\left(\frac{\ell_i+\ell_j}{2}\right)
\,\sinh\!\left(\frac{|\ell_i-\ell_j|}{2}\right),
\]
we obtain
\begin{equation}
P(\ell)
\;\propto\;
\exp\!\left(\frac{p+1+2q}{2}\sum_i \ell_i\right)
\prod_{i<j}\sinh\!\left(\frac{|\ell_i-\ell_j|}{2}\right).
\label{eq:hyperbolic_vandermonde}
\end{equation}

\subsection{Separation into scale and shape}

Define the mean log-eigenvalue $\bar\ell = (1/p) \sum_{i=1}^p \ell_i$, 
and define shape variables $r_i = \ell_{i} -\bar{\ell}$ whence $\sum_i r_i = 0$.

Then $\bar\ell = \frac{1}{p}\log|\bC|$ controls the determinant,
while $r=(r_1,\dots,r_p)$ parametrises each \emph{determinant fibre} (i.e. the set of all matrices with a common determinant).

Because the transformation $\ell \mapsto (\bar\ell,r)$ is linear, the Lebesgue measure
splits as
\begin{equation}
d^p\ell = \sqrt{p}\, d\bar\ell\, d^{p-1}r,
\end{equation}
where $d^{p-1}r$ is the Lebesgue measure on the hyperplane $\sum_i r_i=0$.
We can ignore the constant $\sqrt{p}$, and substituting $\ell_i=\bar\ell+r_i$ into \eqref{eq:hyperbolic_vandermonde} gives
\begin{equation}
P(\bar\ell,r)
\;\propto\;
\exp\!\left[\frac{p(p+1+2q)}{2}\bar\ell\right]
\prod_{i<j}\sinh\!\left(\frac{|r_i-r_j|}{2}\right).
\end{equation}

\subsection{Conditioning on fixed determinant}

Fixing the determinant $|\bC|$ is equivalent to conditioning on $\bar\ell=\mathrm{const}$.
The exponential prefactor then becomes an overall constant and drops out.
The induced density on the fibre is therefore
\begin{equation}
P_{\text{fibre}}(r)
\;\propto\;
\prod_{i<j}\sinh\!\left(\frac{|r_i-r_j|}{2}\right),
\qquad
\sum_i r_i=0.
\label{eq:fibre_density}
\end{equation}

\subsection{Regularisation}

The density \eqref{eq:fibre_density} is improper: it diverges logarithmically for
large spectral spreads.  For numerical sampling we therefore introduce a weak,
rotationally invariant confining term
\begin{equation}
P_\tau(r)
\;\propto\;
\prod_{i<j}\sinh\!\left(\frac{|r_i-r_j|}{2}\right)
\exp\!\left(-\frac{\tau}{2}\|r\|^2\right),
\label{eq:sample}
\end{equation}
with $\tau\ll 1$.
This regularisation ensures normalisability while leaving the posterior dominated
by the likelihood (e.g.\ a Wishart or Student-$t$ factor).

\subsection{Implications for sampling}

Equation \eqref{eq:sample} defines a smooth log-concave target function on the
$(p-1)$-dimensional subspace $\sum_i r_i=0$.  A Hamiltonian sampler operating directly on $r$ 
respects the correct Lebesgue-induced measure, with the correct eigenvalue repulsion, and is combined with an independent draw of $\bar\ell$.

We separately generate random orthogonal matrices $\bQ$ via a standard QR decomposition. A $p\times p$ matrix with independent standard normal entries is QR-decomposed, and the signs of the columns of $\bQ$ were adjusted using the diagonal of R. The resulting matrix is Haar-distributed on $O(p)$ \citep{Stewart1980}.  Combining with the samples of $\lambda_i$, this yields samples of $\bC=\bQ\bLambda \bQ^\top$ with prior density \eqref{eq:sample} above $|\bC|=1$.

\section{Python implementation}
\label{app:code}
\renewcommand{\theequation}{D.\arabic{equation}}
\setcounter{equation}{0}

We provide here a Python function for the moment-matched natural log-likelihood.  Note that we do not provide a Stan code to sample the mixture representation parameter $\tau$ for an accurate log-likelihood evaluation at fixed model parameters, since it will be very slow, with one Stan run per evaluation.  It is much more efficient for Stan to sample the model parameters and $\tau$ jointly.

\begin{verbatim}
import numpy as np
from scipy.linalg import cho_factor, cho_solve
from scipy.special import gammaln

def loglike_student_t_gauss_mm(data, mu, S_hat, C_g, Nsim, method="percival", Ntheta=2):
    """
    Moment-matched Student-t (natural) log-likelihood for convolution of
        Student-t(S_hat, Nsim) and Gaussian(C_g) distributions

    Parameters
    ----------
    data : (p,) array
    mu   : (p,) array, model expectation value of data
    S_hat: (p,p) estimated covariance from simulations
    C_g  : (p,p) additional known Gaussian covariance
    Nsim : int, number of simulations used for the S_hat covariance estimate 
    method : "Percival" (Percival et al 2022; default) or "SH" (Sellentin-Heavens 2016)
    Ntheta : int, number of model parameters

    Returns
    -------
    logL       : float, natural log of likelihood (up to an additive constant)
    nu_star    : float, approximate Student-t degrees of freedom
    Sigma_star : (p,p) array, approximate Student-t scale matrix
    """

    p = data.size
    r = data - mu

    # Base Student-t parameters
    if method.lower() == "sh":
        nu = Nsim - p

    elif method.lower() == "percival":
        B  = (Nsim - p - 2) / ((Nsim - p - 1)*(Nsim - p - 4))
        m  = 2 + Ntheta + (Nsim - 1 + B*(p - Ntheta)) / (1 + B*(p - Ntheta))
        nu = m - p

    else:
        raise ValueError("Method must be 'Percival' or 'SH'")

    if nu <= 4:
        raise ValueError("Need nu > 4 for kurtosis matching.")

    # Base Student-t scale matrix (not covariance)
    Sigma = (Nsim - 1)/nu * S_hat

    # Covariance of the base Student-t
    C_t = (nu/(nu - 2.0)) * Sigma

    # Moment-matched total covariance
    C_star = C_g + C_t

    # Kurtosis matching
    Nmat = np.linalg.solve(C_star, Sigma)
    trN  = np.trace(Nmat)
    trN2 = np.trace(Nmat @ Nmat)

    kappa = (2.0 * nu**2) / ((nu - 2.0)**2 * (nu - 4.0))
    kappa *= (trN**2 + 2.0 * trN2)

    nu_star = 4.0 + 2.0 * p * (p + 2.0) / kappa

    # Effective Student-t scale matrix
    Sigma_star = ((nu_star - 2.0)/nu_star) * C_star

    # Log-likelihood under the moment-matched Student-t
    cho    = cho_factor(Sigma_star, lower=True, check_finite=False)
    x      = cho_solve(cho, r, check_finite=False)
    quad   = float(r @ x)
    logdet = 2.0 * np.sum(np.log(np.diag(cho[0])))

    logL = (
        gammaln(0.5*(nu_star + p)) - gammaln(0.5*nu_star)
        - 0.5*(p*np.log(nu_star*np.pi) + logdet)
        - 0.5*(nu_star + p)*np.log1p(quad/nu_star)
    )

    return float(logL), float(nu_star), Sigma_star
\end{verbatim}

\bibliographystyle{mnras}
\bibliography{references}

\end{document}